# Photonic crystal spatial filtering in broad aperture diode laser

S.Gawali [1,a)], D. Gailevičius [2,3], G. Garre-Werner [1,4], V. Purlys [2,3], C. Cojocaru [1], J. Trull [1]
J. Montiel-Ponsoda [4], and K. Staliunas [1,5]

[1] Universitat Politècnica de Catalunya (UPC), Physics Department, Rambla Sant Nebridi 22, 08222, Terrassa Barcelona, Spain

[2] Vilnius University, Faculty of Physics, Laser Research Center, Saulėtekio al. 10, LT-10223, Vilnius, Lithuania

[3] Femtika LTD, Saulėtekio al. 15, LT-10224, Vilnius, Lithuania

[4] Monocrom S.L, Vilanoveta, 6, 08800, Vilanova i la Geltrú, Spain

[5] Institució Catalana de Recerca i Estudis Avançats (ICREA), Passeig Lluís Companys 23, 08010, Barcelona, Spain

a) Author to whom correspondence should be addressed: sandeep.babu.gawali@upc.edu

ABSTRACT

Broad aperture semiconductor lasers usually suffer from low spatial quality of the emitted beams. Due to the highly compact character of such lasers the use of a conventional intra-cavity spatial filters is problematic. We demonstrate that extremely compact Photonic Crystal spatial filters, incorporated into the laser resonator, can improve the beam spatial quality, and correspondingly, increase the brightness of the emitted radiation. We report the decrease of the $M^2$ from 47 down to 28 due to Photonic Crystal spatial intra-cavity filtering, and the increase of the brightness by a factor of 1.5, giving a proof of principle of intra-cavity Photonic Crystal spatial filtering in broad area semiconductor lasers.

Broad Aperture Semiconductor (BAS) lasers usually suffer from the low spatial quality of the emitted beams, especially in high power emission regimes. This, among others, imposes limitations for tight focusing or coupling of its radiation into the optical fibers. The strong divergence of the beam in the fast-axis ($y$-axis) does not represent a problem since the single-transverse-mode-character of the radiation in this direction allows perfect collimation of the beam. Collimation in the slow-axis ($x$-axis) instead is quite problematic due to the intrinsic multi-transverse-mode emission along this axis. The absence of intra-cavity spatial filtering is the reason of the poor beam quality of other micro-laser types, such as microchip lasers, or Vertical Cavity Surface Emission Lasers (VCSELs). Conventional lasers (eg. solid state lasers) usually use intra-cavity spatial filters, typically using a confocal arrangement of lenses with a diaphragm in the confocal plane. This configuration allows direct access to the far field, where the diaphragm acts as a "low-pass" spatial filter blocking the higher angular components. This type of spatial filter has been demonstrated in an external resonator of diode laser array to improve the beam quality [1]. However, such a filtering design is very inconvenient or even impossible for intra-cavity use in micro-lasers, such as microchip or semiconductor lasers, or VCSELs, since the lengths of the resonators typically lies in the millimeter range (in VCSELs case even in micrometer range) and provide no space for direct access to the far field. For the case of high power diode lasers, usually configured in an array of multiple emitters, the use of bulky lenses is impossible.

The quality of the BAS laser beams at low powers can be improved by several methods, for instance: (i) confining transversally the radiation by inscribing waveguide structure in the semiconductor material. This microstructured fabrication, however, severely restricts the amplification area, and thus reduces the emission power; (ii) by restricting the amplification/emission area through the use of apertures in VCSELs; or (iii) by accurate gain guiding in microchip lasers. None of these methods are useful to solve the problem of beam quality in high-emission-power regimes. For the BAS lasers, some other techniques have been proposed, like the use of tapered geometries, evanescent spatial filtering, and the use of external cavity [2-4]. However, these methods have a limitation in terms of power achievable and the size of the device.

A promising idea to solve the beam quality problems of the microlasers in the high emission power regime is the use of intra-cavity Photonic Crystal (PhC) spatial filters integrated in the laser cavity. The concept of spatial filtering using PhCs was proposed in [5, 6], and subsequently experimentally demonstrated in [7-9]. The idea is based on a selective deflection of the angular components of the light propagating through a 2-D photonic structure: the angular components of the incident light resonant with the transverse and longitudinal periodicities of the photonic structure diffract efficiently and are deflected from the zero-diffraction order of the transmitted beam. The use of such PhC spatial filtering [10] was already proved as a powerful tool to clean the spatial structure of the beams, which could be efficiently employed for instance in the microchip lasers [11]. For more detailed theory on PhC spatial filtering see review [12].

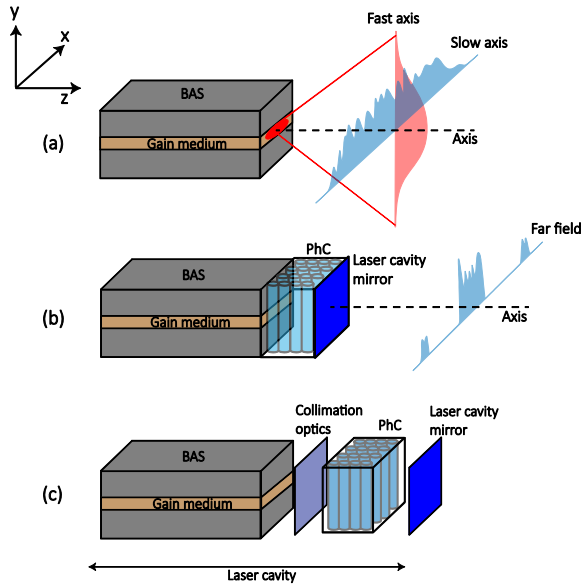

**Fig. 1.** (a) The illustration of the irregular beam structure emitted by typical BAS lasers (with partially reflective output facet mirror) in high power regimes. (b) The idea of spatial filtering by monolithically integrated PhCs in a compact configuration. While the lower angle modes propagate unaffected, as shown in a schematic far-field profile, the PhC diffracts and eliminates the higher angle modes. (c) The PhC placed inside an extended cavity resonator, which mimics the situation of the compact intra-cavity spatial filtering shown in (b).

The ultimate goal of the research reported in this article is the design of efficient intra-cavity PhC acting as a spatial filter in BAS lasers. The possible monolithic implementation scheme is illustrated in Fig. 1(b), showing how the PhC structure could be integrated directly between the active medium and the laser output cavity mirror. Such integration is, however, very challenging technologically. In order to show the reliability of the physical principle, i.e. to test the spatial filtering effect of the PhC, we propose in this article an extended-cavity configuration designed to mimic the action of the more compact cavity. A low antireflective (AR) coating at the output facet of the semiconductor material prevents the lasing in the semiconductor gain medium alone, and enables building of the extended resonator by using external mirrors. Collimation optics and an external output cavity mirror provides the necessary feedback to achieve the laser action in this extended-cavity configuration, which allows placing the PhCs inside the laser cavity to test their filtering capabilities, as schematically depicted in Fig. 1(c).

The effect of the PhC filtering in BAS lasers has been numerically studied in extended cavity configuration [13]. In this article, we report the successful proof of the physical implementation of this idea, showing how the sub-millimeter length PhCs affects the emission of the BAS laser. The intra-cavity use of such filters allows an improvement of the beam spatial quality and an increase of brightness of the emitted radiation in high-power regimes. These results open the way for promising future goal i.e. the technological implementation of a compact (monolithic) intra cavity designed system for spatial filtering. This task is beyond the scope of the current letter due to technological challenges.

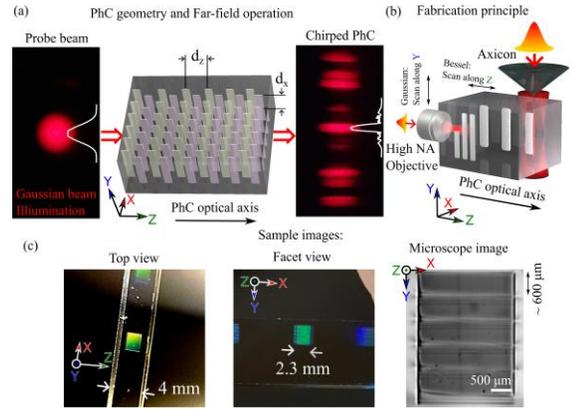

**Fig. 2.** (a) Illustration of the PhC, showing the architecture of index modulation inside the glass substrate and the expected effect on the Far-field intensity spectra of the beam, (b) Illustration of the PhC fabrication using femtosecond pulsed Bessel beam, and (c) photographs and microscope image of the fabricated structures. The images of the Far-field beam intensity profiles show the principle of angular filtering resulting in narrow angular transmission bands.

The PhC samples having a 2D periodic refractive index modulation (Fig. 2(a)) were fabricated via femtosecond laser writing technique in the bulk of glass substrate. The substrates were made from N-BK7 ($n_{ref} \approx 1.52$) glass with broadband antireflective coating on both facets. We fabricated two versions of PhC filters: the ones inscribed by Gaussian beam, and others inscribed using Bessel beam (Fig. 2(b)). For more details about the fabrication procedures and exposure parameters see [11] for Gaussian beam fabrication. The Gaussian beam fabrication strategy allows flexible 3D geometry but is limited by the working distance of high numerical aperture (NA) focusing (microscope) lens, therefore the maximal length of the PhC in our case was limited to < 375 μm. In addition, a significant amount of spherical aberration is present, which leads to defects and distortions of the PhC geometry. Moreover, the spatial filtering performance (the angular range of deflected components) is proportional to the length of the PhC $\Delta\alpha = \Delta n\, l/\lambda$, where $\Delta n$ is the change in refractive index, which is of the order of $10^{-3}$. Due to the limited length of PhCs the Gaussian beam fabrication strategy allowed a filtering range of ~1 degree, which was not wide enough for BAS lasers (~ 5-10 degrees preferred).

The Bessel beam fabrication of PhCs [14] has practically no limits of longitudinal periods, due to the fact that the PhCs are fabricated by a beam illuminating along the perpendicular $y$-direction. Using this technique photonic structures of millimeters or even centimeters in length can be fabricated [15]. By scanning the sample as shown in Fig. 2(b), long (~ 600 μm) refractive index modified bulk structures were produced. We achieved this by taking a UVFS axicon with an apex angle of 179° and illuminating it with a collimated Gaussian intensity profile pulsed (200 fs) laser beam with a diameter of $2w = 5.3$ mm $(1/e^2)$, 1030 nm wavelength, pulse repetition rate of 25 kHz and 8 μJ pulse energy. We demagnified the resulting Bessel beam with a ~55× telescope producing a 600 μm length Bessel zone inside the substrate. Modified index areas were around half the $d_x$ value wide and half of the minimum $d_{zmin}$ value long (we scanned the substrate at a $v = 2500$ μm/s linear velocity while opening and closing the laser shutter to produce modifications of such length). The refractive index change is difficult to measure for such a closely packed structure, however we expect it to be near $\Delta n \lesssim 3 \cdot 10^{-3}$ [16, 17].

The geometry of the structures is characterized by the transverse and longitudinal lattice constants $d_x$ and $d_z$. In order to increase the angular width of the filtering area, the longitudinal period can be "chirped", i.e. constantly varied along the propagation direction, as also described in [9,14]. The chirped structures are additionally characterized by the values of the longitudinal periods corresponding to the first, $d_{z1}$, and the last, $d_{z2}$, periods, with $d_z$ values changing linearly along the structure. Following the previous references [12] we use the geometry factor $Q$ to characterize our PhC structures. This parameter defines a relation between the transverse and longitudinal periods of the structure (not to be confused with resonator finess factor) and is defined as $Q = 2d_x^2 n_{ref}/\lambda d_z$, and can be used to estimate the filtering angle $\sin\alpha \approx \lambda(Q-1)/2d_x n_{ref}$. Small variation of the geometry factor result in a linear change of the central value of the filtering angle $\Delta Q \sim \Delta \alpha$.

Following this notation, the Gaussian-beam fabricated (non-chirped) structure was characterized by $d_x = 2$ μm, $\lambda = 970$ nm and $Q = 0.9$ with an aperture of 2x2 mm$^2$ and number of longitudinal periods $N_L = 10$. The Bessel-beam fabricated structure had a different transverse period $d_x = 3$ μm and was chirped along the $z$ direction in the range of $1.15 \leq Q \leq 1.27$ with $N_L = 80$ and an aperture of 2.85 x 2.3 mm$^2$. In order to increase the height of the PhC aperture 5 layers of modified regions were stitched vertically. An example of this structure is shown in Fig. 2(c).

The filtering performance of the fabricated PhCs can be described through their transfer function, i.e. their diffracting action on the spatial modes, although for their proper operation these filters should be placed in the near field plane. The transmittance of the fabricated filters is shown in Fig. 2(a). When inserted inside the laser cavity, the diffracted modes do not contribute to the lasing action modifying the overall output of the system and the dynamics of the lasing operation.

For the experiments we used a BAS laser of 400 μm transverse width (along $x$-axis) and 1500 μm length (along $z$-axis), emitting 1.3 W in CW regime when operated at 3A at a wavelength of 970 nm. In the experiment we drove the laser at a repetition rate of 50 Hz with a 25% duty cycle. To operate this laser in the extended-cavity configuration the output facet of the laser was coated with a low AR coating ($R<0.01\%$) to prevent the emission of any optical mode above the lasing threshold level from the gain medium alone and a 4% reflectivity mirror was used as the output cavity laser mirror. The reflectivity of this external mirror is limited to 4% to avoid the possible damage of the emitter facets [18]. The extended cavity configuration, shown schematically in Fig. 3, was implemented using cylindrical lenses acting separately in the slow and fast axis. The fast axis emission was collimated by mounting a cylindrical lens of very short focal length of 590 μm, with a NA of 0.8. (FAC lens in Fig. 3). For the slow-axis direction, a double 4-f system was used to obtain two conjugated planes of the output facet of the active medium. In this way, the PhC could be inserted in the first conjugated plane, with 1:1 magnification, designated as B in Fig. 3. The PhC filters were mounted on a 3D translational stage, which allowed inserting it properly into the beam path. The size of the PhC 2x2 mm$^2$ for Gaussian beam fabricated crystal and 2.85x2.3 mm$^2$ for Bessel beam fabricated crystal, was large enough to allow the entire beam pass through the filter. The second 4f system provided a second near field plane, marked as D in Fig. 3, where the output laser-cavity mirror was placed and carefully aligned to obtain lasing action. All lenses were AR coated in order to prevent the appearance of multiple cavities effect within the setup.

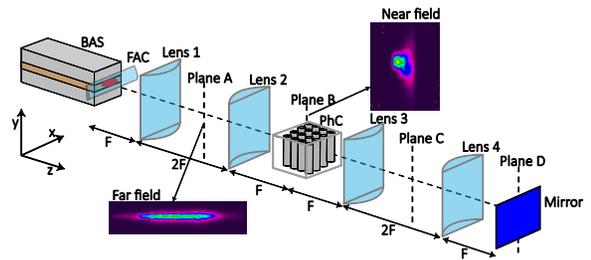

**Fig. 3.** Experimental setup of spatial filtering in extended-cavity configuration. BAS: Broad Area Semiconductor, FAC: fast-axis collimator. Lenses 1-4 with a focal length of 50.8 mm. Far-field areas are marked as (plane A) and (plane C). Near-field planes (images of the BAS output facets marked as B and D). The reflectivity of the mirror is 4%.

The emission characteristics of this extended laser when no PhC is inserted in the cavity shows a threshold at 1.3 A, and a spectral emission centered at 970 nm with a spectral bandwidth of 2.9 nm. The beam profile at the output of the laser (plane D in Fig. 3) was recorded by imaging this plane into a CCD camera (Spiricon SP620U) with proper magnification. The near-field profile recorded by this means has dimensions of 450 μm x 1.4 mm. The effect of the PhC filtering was observed experimentally by recording the far-field distribution of the emitted radiation. The far-field pattern in our experiments was recorded using an external lens out of the cavity (not shown in Fig. 3) to image the rear focal plane into the CCD.

Fig. 4(a) shows the far-field profile of the laser emission when no filter and when the Gaussian-fabricated and Bessel-fabricated PhCs are placed within the laser cavity. The PhC effectively filters the beam along the slow-axis by selectively deflecting the higher-order modes in a given frequency range. The spectral selectivity of the filters is better seen by plotting the ratio between the far-field profile without and with filters inserted in the lasing cavity. This ratio, plotted on Fig. 4(b), indicates a clear filtering band between 1.5 and 2.5 degrees, with a more pronounced effect of the Bessel-fabricated PhC.

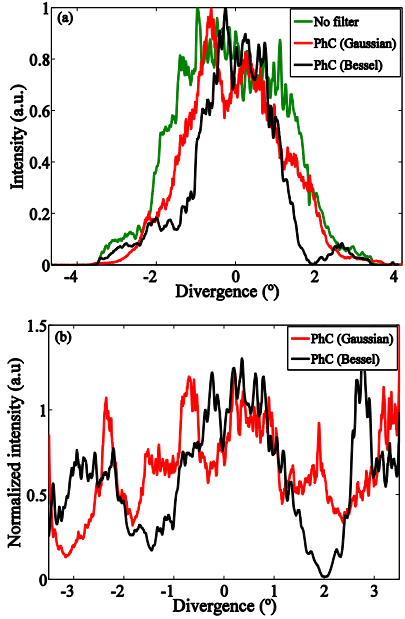

**Fig. 4.** (a) Measured Far-field profile without PhC (green), with PhC Gaussian beam (red) and Bessel beam fabricated (black) technique. (b) Reconstructed ratio between the far-field profiles in (a) without and with filters inserted in the lasing cavity.

The spatial quality of the beam is quantified by the measurement of the beam quality factor and brightness. The first is given by $M^2 \approx BPP_{actual\ beam}/BPP_{Gaussian\ beam}$, where BPP refers to beam parameter product and is given by $BPP = Beam\ waist \times Divergence\ angle$. The brightness, $B$ is defined as $B = P_{opt}/\lambda^2 M_{slow}^2 M_{fast}^2$ [19], where $P_{opt}$ is the average optical power, $M_{slow}^2$ and $M_{fast}^2$ are the beam quality factors along slow and fast axis respectively and $\lambda$ is the central wavelength.

In our setup, the $M^2$ factor was calculated by recording the profile of the beam as a function of propagation distance after being focused using an ancillary external lens of 100 mm focal length. The beam diameter, recorded independently in the fast and slow axis, was determined using the D4σ method (corresponding to 4 times the standard deviation of the energy distribution evaluated in the transverse direction over the beam intensity profile) [20]. From the diameter versus distance plot, the minimum spot size and divergence angle was measured to determine $M^2$. The initial $M^2$ factor measured from our laser with no filtering action was of $M_{slow}^2 = 47$ and $M_{fast}^2 = 3.3$. These large values obtained in the slow-axis ($x$-direction) indicate clearly the poor quality of the beam and are consistent with reported values in similar systems [21]. The unexpected large value of $M^2$ in the fast-axis ($y$-direction) is due to a small misalignment of the FAC lens in this particular sample available.

The spatial filtering effect of the PhCs in a single transmission when placed outside of the laser cavity has been proved in previous publications [8, 9]. However, this linear single-pass action, while improving the spatial quality of the beam (reducing $M_{slow}^2$), does not increase the brightness of the emitted radiation. Brightness enhancement is possible by intra cavity spatial filtering only. For instance, using conventional confocal lens arrangements, the suppression of the higher order transverse modes concentrates most of the pumping energy into the lowest order modes. The increase of brightness is obtained because closing the aperture has a weak effect on the total intensity, which decreases weakly, but a strong effect on the beam divergence, which can decrease considerably. If the aperture is too narrow, starting to affect the lowest transverse modes, the brightness starts decreasing as well.

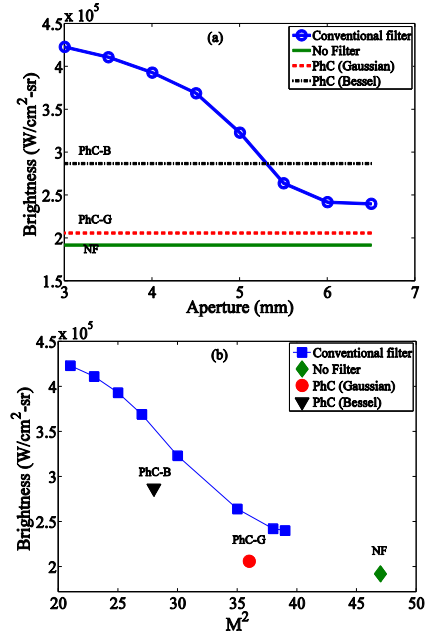

**Fig. 5.** (a) Shows the brightness of emission as a function of aperture (blue line) compared with PhC fabricated by Gaussian (PhC-G) and Bessel beam (PhC-B) and with No filter (NF). (b) Brightness versus $M^2$ along the slow axis.

Before exploring the effect of our fabricated PhC filters, we checked the spatial filtering properties using a variable thickness slit placed at the far field plane inside the extended laser resonator (plane C shown in Fig. 3). This situation allows us to compare the action of the filters with respect to a more typical configuration. The initial value of $M^2$ and brightness of our laser without any type of filtering is shown as the green diamond in Fig. 5(b). The results of the brightness as a function of the aperture (width) of the slit are shown in Fig. 5(a). The horizontal green line indicates the value of the brightness for the case when no aperture was placed in the cavity. The calculated brightness as a function of aperture width, shows that the decrease in divergence is accompanied by brightness increase by a factor of almost 2.2.

While the aperture acts on the transverse modes when placed in the far-field plane, the PhC spatial filters work when placed inside the cavity at the near-field plane marked as B. This is achieved with the help of 4-f system configuration as shown in Fig. 3. This is a fundamental difference between the conventional far-field filtering, and the PhC near field filtering, since the filtering with the PhC, positioned in the near field domain, is the only option that could be implemented in a monolithic system. The brightness of the laser with PhCs intracavity filters are indicated by the horizontal (dashed and dashed-dot) lines in Fig. 5(a).

The measurements with the PhCs fabricated using the Gaussian beam technique show that the $M^2_{slow}$ factor decreased from 47 to 36. As the output optical power dropped from 0.31 W to 0.230 W, the reduction of the divergence was not sufficient to compensate the loss of power and hence the brightness did not increase as indicated in Fig. 5(b) (red circle). Next, we explored the Bessel beam fabricated crystals. A clear modification of the emission pattern is observed (Fig. 4(a)). The change is observable with a slight loss in power. For the Bessel-beam fabricated PhCs, the $M^2_{slow}$ reduced from 47 to 28 while brightness increased by a factor of 1.5 (black triangle in Fig. 5(b)).

The results of spatial filtering for conventional confocal filtering, PhC (Gaussian beam) and PhC (Bessel beam) compared with the situation without filtering are summarized in Fig. 5(b). Although the best performance was obtained for the conventional confocal filtering case, it is convincingly shown that the action of the improved PhCs filters tends towards the achievement of these optimal values.

Concluding, we have demonstrated spatial filtering in broad area semiconductor laser using a PhC spatial filter in extended cavity configuration, which mimics the compact cavity configuration. The main result was a decrease of $M^2_{slow}$ and the brightness enhancement of the emitted radiation. Comparison with the conventional confocal filtering technique shows that by using PhCs filtering, one can achieve results approaching the same values of spatial beam quality improvement. This is the first test to demonstrate the working principle of spatial filtering in Broad Area Laser with single emitter with output power of the order of 1W. The same technique can be applied to high power diode laser bars. The advantage of the use of PhCs filtering comparing with the conventional technique is that it offers the possibility of integration of the PhC into the BAS leading to a serious breakthrough in regimes of high power emission. This article does not demonstrate such integration due to limited technologies, however, it shows the potential for such integration in near future, by proving the physical principles.

This work was supported by the EUROSTARS Project E-10524 HIP-Lasers, as well as by Spanish Ministerio de Ciencia e Innovación, and European Union FEDER through project FIS2015-65998-C2-1-P. D.G. and V.P. acknowledge the financial support from "FOKRILAS" (Project S-MIP-17-109) from Research Council of Lithuania.